\documentclass[aps,prl,twocolumn,superscriptaddress,footinbib]{revtex4-2} 

\usepackage{graphicx}  
\usepackage{subfigure}
\usepackage{dcolumn} 
\usepackage{bm}
\usepackage{amssymb}   
\usepackage{amsfonts}   
\usepackage{comment}
\usepackage[colorlinks,linkcolor=blue,anchorcolor=blue,citecolor=blue,urlcolor=blue]{hyperref}
\usepackage{xcolor}
\usepackage{soul}
\usepackage{tabularx,booktabs,ragged2e,multirow}
\newcolumntype{C}{>{\centering\arraybackslash}X}
\usepackage{multirow}
\usepackage{amsmath}
\usepackage{mathrsfs}
\usepackage{mathcomp}
\usepackage{textcomp}
\usepackage{dsfont}
\usepackage{esint}
\usepackage{braket}
\usepackage{textgreek}
\usepackage{lipsum}
\usepackage{marvosym} 
\usepackage{centernot}
\usepackage{cancel}
\usepackage{dashrule}
\usepackage[normalem]{ulem}

\begin{document}

\preprint{APS/123-QED}

\title{First-Order $\mathcal{PT}$ Phase Transition in Non-Hermitian Superconductors}

\author{Xuezhu Liu}
\affiliation{Center for Advanced Quantum Studies, School of Physics and Astronomy, Beijing Normal University, Beijing 100875, China}
\affiliation{Key Laboratory of Multiscale Spin Physics (Ministry of Education), Beijing Normal University, Beijing 100875, China}

\author{Ming Lu}
\email{Correspondence: luming@baqis.ac.cn}
\affiliation{Beijing Academy of Quantum Information Sciences, Beijing 100193, China}

\author{Haiwen Liu}
\email{Correspondence: haiwen.liu@bnu.edu.cn}
\affiliation{Center for Advanced Quantum Studies, School of Physics and Astronomy, Beijing Normal University, Beijing 100875, China}
\affiliation{Key Laboratory of Multiscale Spin Physics (Ministry of Education), Beijing Normal University, Beijing 100875, China}
\affiliation{Interdisciplinary Center for Theoretical Physics and Information Sciences, Fudan University, Shanghai 200433, China}

\author{X. C. Xie}
\affiliation{Interdisciplinary Center for Theoretical Physics and Information Sciences, Fudan University, Shanghai 200433, China}
\affiliation{International Center for Quantum Materials, School of Physics, Peking University, Beijing 100871, China}

\newcommand{\br}{{\bm r}}
\newcommand{\bk}{{\bm k}}
\newcommand{\bq}{{\bm q}}
\newcommand{\bp}{{\bm p}}
\newcommand{\bv}{{\bm v}}
\newcommand{\bmm}{{\bm m}}
\newcommand{\bA}{{\bm A}}
\newcommand{\bE}{{\bm E}}
\newcommand{\bB}{{\bm B}}
\newcommand{\bH}{{\bm H}}
\newcommand{\bd}{{\bm d}}
\newcommand{\bzero}{{\bm 0}}
\newcommand{\bOmega}{{\bm \Omega}}
\newcommand{\bsigma}{{\bm \sigma}}
\newcommand{\bJ}{{\bm J}}
\newcommand{\bL}{{\bm L}}
\newcommand{\bS}{{\bm S}}
\newcommand\dd{\mathrm{d}}
\newcommand\ii{\mathrm{i}}
\newcommand\ee{\mathrm{e}}
\newcommand\zz{\mathtt{z}}
\makeatletter
\let\newtitle\@title
\let\newauthor\@author
\def\ExtendSymbol#1#2#3#4#5{\ext@arrow 0099{\arrowfill@#1#2#3}{#4}{#5}}
\newcommand\LongEqual[2][]{\ExtendSymbol{=}{=}{=}{#1}{#2}}
\newcommand\LongArrow[2][]{\ExtendSymbol{-}{-}{\rightarrow}{#1}{#2}}
\newcommand{\cev}[1]{\reflectbox{\ensuremath{\vec{\reflectbox{\ensuremath{#1}}}}}}
\newcommand{\red}[1]{\textcolor{red}{#1}} 
\newcommand{\blue}[1]{\textcolor{blue}{#1}} 
\newcommand{\green}[1]{\textcolor{orange}{#1}} 
\newcommand{\mytitle}[1]{\textcolor{orange}{\textit{#1}}}
\newcommand{\mycomment}[1]{} 
\newcommand{\note}[1]{ \textbf{\color{blue}#1}}
\newcommand{\warn}[1]{ \textbf{\color{red}#1}}

\makeatother

\begin{abstract}
The interplay between superconductivity and environmental dissipation, effectively captured by non-Hermitian Hamiltonian, is a new frontier for exotic quantum phases. We explore a $\mathcal{PT}$-symmetric non-Hermitian superconductor with balanced gain and loss. To ensure experimental relevance, we develop a right-eigenstate-based non-Hermitian  mean-field theory. We uncover a novel first-order phase transition that coincides exactly with the $\mathcal{PT}$ symmetry breaking point, driven by the interplay between superconducting pairing interactions and non-Hermitian dissipation. In the $\mathcal{PT}$-symmetric phase, moderate NH dissipation enhances superconductivity, while in the $\mathcal{PT}$-broken phase, intensified dissipation significantly suppresses it. These phenomena, characterized by abrupt jumps in observables, can be probed through localized spectral measurements and macroscopic superfluid density analysis. Additionally, the stability analysis offers robust theoretical insights to support experimental investigations of this unique transition.
\end{abstract}


\maketitle

\mytitle{Introduction}.--The exploration of novel quantum states is a central theme in condensed matter physics. While unconventional superconductors \cite{RevModPhys.63.239, RevModPhys.78.17, Stewart1984, RevModPhys.83.1589, PhysRev.135.A550, larkin1965inhomogeneous} and non-equilibrium superconductors \cite{kopnin2001theory,Fausti2011,Kundu2013,Mankowsky2014,Khim2021} have been widely studied, most studies focus on closed quantum systems governed by Hermitian Hamiltonians. In reality, however, superconductors inevitably couple to environments, resulting in dissipation and nonunitary dynamics. This open quantum behavior can be effectively captured by non-Hermitian (NH)  Hamiltonians \cite{moiseyev2011non}, which have uncovered a range of unconventional phenomena in superconducting systems, including nonorthogonal Majorana zero modes \cite{Kawabata2018}, exceptional odd-frequency pairing \cite{Cayao2022}, paramagnetic Meissner effect \cite{Tamura2025} and unusual phase transitions in fermionic superfluids \cite{Yamamoto2019,Yamamoto2021}.

\begin{figure}[t]
\includegraphics[width=8.6cm]{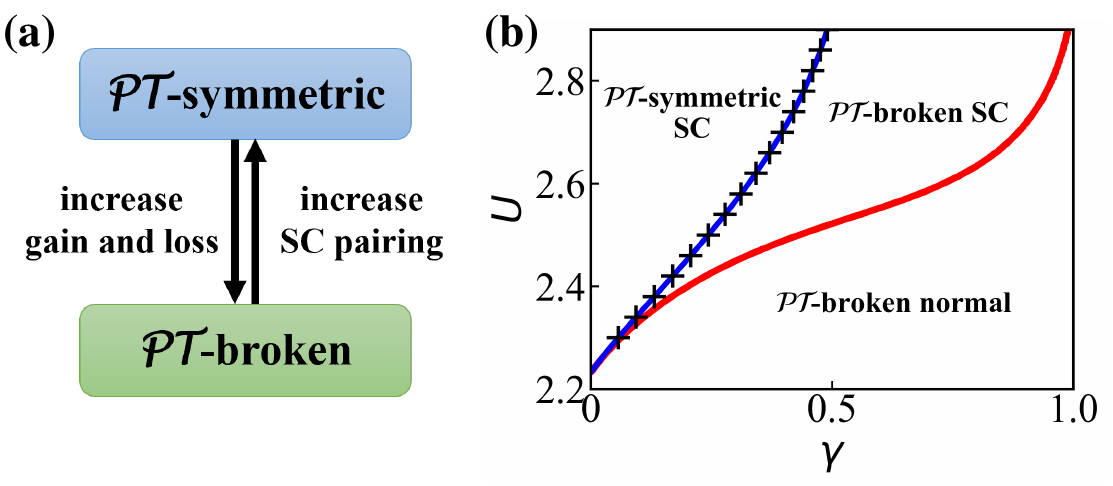}
\caption{(a) The contrasting effects of superconducting pairing interactions and gain and loss on $\mathcal{PT}$ symmetry. (b) $\mathcal{PT}$ phase diagram at half-filling. The black crosses mark the points where the order parameter exhibits jumps. The blue curve indicates the onset of complex eigenenergies, while the red curve corresponds to the onset of $\Delta_0 = 0$.}
\label{fig:fig1}
\end{figure}

Within the broader class of NH systems, significant focus has been placed on those exhibiting \(\mathcal{PT}\) symmetry, as it allows for real energy spectrum in the \(\mathcal{PT}\)-symmetric phase \cite{Bender1998a,Mostafazadeh2002,Mostafazadeh2002a,Mostafazadeh2002b}. Here, \(\mathcal{P}\) and \(\mathcal{T}\) denote parity and time-reversal operations, respectively. Balanced gain and loss are a common mechanism for realizing \(\mathcal{PT}\) symmetry, and have been realized in various physical platforms \cite{bender2019pt,Bender2013,Dembowski2001,Brandstetter2014,Lee2009,Li2019,Regensburger2012,Rueter2010,Wu2019,Xu2016,Yu2024}. Although \(\mathcal{PT}\) symmetry has been widely explored in single-particle and optical setups, its role in NH superconductors remains relatively unexplored \cite{Chtchelkatchev2012,Kornich2022,Kornich2022a}, particularly in understanding the interplay between \(\mathcal{PT}\) symmetry and Cooper pairing. While recent studies have begun to explore PT-symmetric superconductors, the definition of order parameter remained ambiguous, and the nature of the phase transition driven by the competition between pairing and dissipation was uncharted \cite{Ghatak2018}.

In this work, we explore a $\mathcal{PT}$-symmetric NH superconducting ($\mathcal{PT}$NHS) system on a honeycomb lattice with balanced gain and loss. We reveal that $\mathcal{PT}$ symmetry introduces a novel first-order superconducting $\mathcal{PT}$ phase transition. This stands in stark contrast to the conventional understanding that superconducting transitions are typically second-order, except for the rare cases like FFLO states where the order parameter is spatially inhomogeneous \cite{PhysRev.135.A550,larkin1965inhomogeneous}, suggesting a fundamentally new mechanism driven by dissipation. The transition can be detected through localized spectral measurements and macroscopic superfluid density analysis \cite{Scalapino1993}. 

As illustrated in Fig.~\ref{fig:fig1}(a), gain and loss break $\mathcal{PT}$ symmetry, whereas the superconducting pairing interaction preserve it. The interplay between these contrasting effects governs the superconducting $\mathcal{PT}$ phase transitions. The first-order nature of the transition is characterized by a discontinuous jump in the order parameter and a sudden change in the superconducting response, as shown in Fig.~\ref{fig:fig1}(b). Meanwhile, moderate dissipation enhances superconductivity, whereas stronger dissipation abruptly suppresses it. The stability of the superconducting phase is affirmed by the condensation-energy analysis, which furnishes a solid theoretical basis for experimental studies of the superconducting $\mathcal{PT}$ transition.

\mytitle{Right-eigenstate-based NH mean-field theory}.--We consider a $\mathcal{PT}$NHS model on honeycomb lattice with s-wave paring interaction \cite{PhysRev.106.162} described by
\begin{equation}
\label{equ:A1:eqdH0}
\begin{aligned}
H= & -w \sum_{\langle n, m\rangle \sigma} c_{n \sigma}^{\dagger} c_{m \sigma}-\sum_{n \sigma} (\mu + i s_n \gamma)  c_{n \sigma}^{\dagger} c_{n \sigma} \\
& -U \sum_n c_{n \uparrow}^{\dagger}  c_{n \downarrow}^{\dagger} c_{n \downarrow} c_{n \uparrow},
\end{aligned}
\end{equation}
where $U$ is pairing interaction strength and $\gamma$ describes the balanced gain and loss on the $a$ ($s_n=1$) and $b$ ($s_n=-1$) sublattices. We fixed $w=1$ as the energy scale and $\mu =0$ (half-filling) in the following.  We emphasis that this Hamiltonian is applicable to a broader range of systems. For example, experimentally more accessible is a purely dissipative system where the dissipation on the $a$ and $b$ sublattices are just unequal. One can always shift a global imaginary energy to obtain the Hamiltonian in Eq.~\eqref{equ:A1:eqdH0}, without affecting the state vectors. According to the Lindblad quantum master equation, this Hamiltonian effectively describes one-body loss originating from the coupling between the environment and superconductors \cite{PhysRevLett.68.580, Suppl}.

The definition of observables plays a central role in quantum mechanics. In NH physics, the eigenstates of a NH Hamiltonian $H$ are defined as $H|\psi^{R}_{i}\rangle=E_{i}|\psi^{R}_{i}\rangle$, which are not orthogonal to each other, i.e., $\langle\psi^{R}_{i} \mid \psi^{R}_{j}\rangle \ne \delta_{ij}$ \cite{Brody2013}.
At the same time, the evolution of states is nonunitary \cite{Suppl}. As a result, the definition of observables requires deliberate choice. Currently, the right-eigenstate-based definition of observables has gradually gained more attention \cite{Meden2023, Sticlet2022, Yamamoto2022, Yan2024, Dora2019, Graefe2008, Yamamoto2021}. The expectation value of operator $O$ for the state $|\psi^{R}\rangle$ is given by $\langle O\rangle={\langle\psi^{R}| O|\psi^{R}\rangle}/{\langle\psi^{R} \mid \psi^{R}\rangle}$, which provides a proper probabilistic interpretation and is experimentally more relevant \cite{Barontini2013,Li2019,Wu2019}.
For example, study on normal metal-insulator-$ \mathcal{PT}$NHS junction have demonstrated that, compared to the biorthogonal definition, the right-eigenstate-based definition correctly captures the Andreev reflection process \cite{Kornich2023}. 

In the right-eigenstate-based NH mean-field theory \cite{Graefe2008, Yamamoto2021}, the order parameter is defined as
\begin{equation}
\begin{aligned}
\label{equ:A1:eqd0}
\Delta_c =  \frac{U}{N} \sum_k \left\langle c_{-k \downarrow} c_{k \uparrow}\right\rangle \equiv \frac{U}{N} \sum_k \frac{\left\langle \Psi_0 \left|c_{-k \downarrow} c_{k \uparrow}\right| \Psi_0\right\rangle}{\langle \Psi_0 \mid \Psi_0\rangle},
\end{aligned}
\end{equation}
where \(N\) is the number of unit cells, \(|\Psi_0\rangle\) is the NH BCS ground state as detailed in below.
Substituting \(c_{-k \downarrow} c_{k \uparrow}=\left\langle c_{-k \downarrow} c_{k \uparrow}\right\rangle+\delta\left(c_{-k \downarrow} c_{k \uparrow}\right)\) into the pairing interaction term and neglecting the second-order terms of $\delta$, we obtain the NH mean-field Hamiltonian:
\begin{align}
H_{MF}  =&-\sum_{k \sigma}\left(w_k a_{k \sigma}^{\dagger} b_{k \sigma} + \Delta_a a_{k \uparrow}^{\dagger} a_{-k \downarrow}^{\dagger}+\Delta_b b_{k \uparrow}^{\dagger} b_{-k \downarrow}^{\dagger}\right) \nonumber \\
  &\quad +\text {h.c.} - i \gamma \sum_{k \sigma}  (a_{k \sigma}^{\dagger} a_{k \sigma}-b_{k \sigma}^{\dagger} b_{k \sigma}) \label{equ:A1:eqd12},
\end{align}
where \(w_k=w\left[e^{-i k_y}+e^{i(\frac{\sqrt{3}}{2}  k_x + \frac{1}{2} k_y)}+e^{i(-\frac{\sqrt{3}}{2}  k_x+\frac{1}{2} k_y)}\right]\).
For the Hermitian case, the inversion symmetry between the \(a\) and \(b\) sublattices ensures \(\Delta_a = \Delta_b\). In the current case, considering the U(1) symmetry of the system's Hamiltonian [Eq.~\eqref{equ:A1:eqdH0}] and the alternating gain and loss on the sublattices, we can generally define
\begin{equation}
\label{equ:A1:eqd255}
\begin{aligned}
\Delta_a=\Delta_0 e^{i \phi}, \quad \Delta_b=\Delta_0^* e^{i \phi},
\end{aligned}
\end{equation}
where \(\phi\) is the U(1) phase and \(\Delta_0 \in \mathbb{C}\). 
Substituting it into Eq.~\eqref{equ:A1:eqd12}, we can diagonalize the NH mean-field Hamiltonian \(H_{MF}=\sum_{\alpha,k,\sigma}E_{\alpha k }\bar{\gamma}_{\alpha k \sigma} \gamma_{\alpha k \sigma} - \sum_{\alpha,k}E_{\alpha k}\),
where \(\alpha = 1, 2\) labels the branches of quasiparticles, and \(\bar{\gamma}_{\alpha k\sigma}, \gamma_{\alpha k\sigma}\) are the quasiparticle operators whose explicit forms are given by Eq.~\eqref{equ:A1:eqs12} in the Supplemental Materials \cite{Suppl}. It is worth pointing out that although \(\bar{\gamma}_{\alpha k\sigma}\) and \(\gamma_{\alpha k\sigma}\) satisfy \(\left \{\bar{\gamma}_{\alpha k\sigma},\gamma_{\alpha'k'\sigma'}\right \} = \delta_{\alpha \alpha'}\delta_{kk'}\delta_{\sigma\sigma'}\), these quasiparticles differ from conventional fermions and bosons due to \(\bar{\gamma}_{\alpha k\sigma} \neq \gamma_{\alpha k\sigma}^\dagger\).
The energy of quasiparticles reads \(E_{\alpha k}= \sqrt {{{\left| {\left| w_k \right| - i (-1)^\alpha \Delta_0 } \right|}^2} - \gamma ^2}\). Thus, \(\text{Re} \Delta_0 \ge \gamma\) ensures that \(\mathcal{PT}\) symmetry is preserved, while \(\text{Re} \Delta_0 < \gamma\) indicates that \(\mathcal{PT}\) symmetry is broken.

\begin{figure}[t] 
\includegraphics[width=8.6cm]{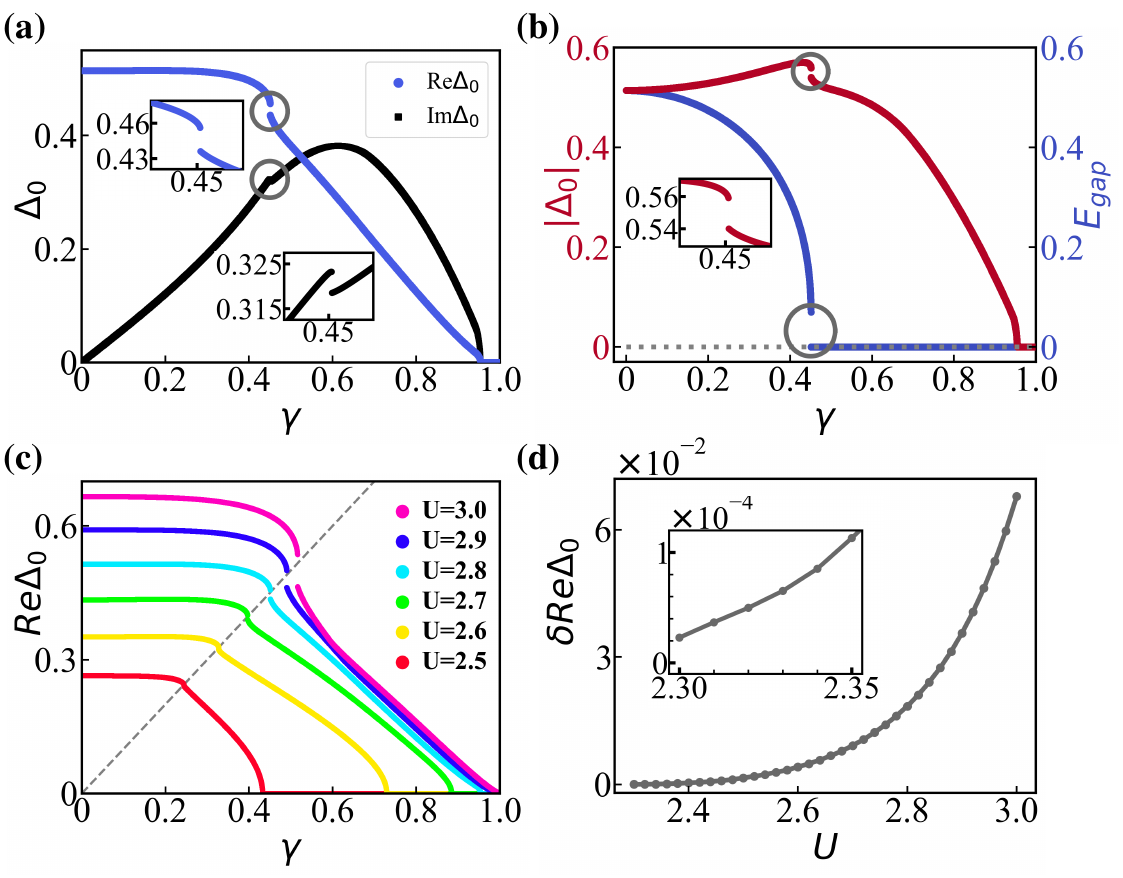}
\caption{Order parameter and spectrum gap as functions of $\gamma$ for $U=2.8$: (a) $\Delta_0$, (b) $|\Delta_0|$ (red) and $E_{\text{gap}}$ (blue). Discontinuous jumps are marked by gray circles, with enlarged views provided in the insets. (c) $\text{Re}\Delta_0$ plotted against $\gamma$ for various values of $U$, where the gray dashed line denotes $\text{Re}\Delta_0 = \gamma$. (d) The jump magnitude $\delta \text{Re} \Delta_0$ plotted against $U$, with an inset highlighting the behavior for small $U$.}
\label{fig:fig2}
\end{figure}

In the ground state, all quasiparticle excitations are absent.
Consequently, $|\Psi_0\rangle = \prod_{\alpha,k} \gamma_{\alpha k \uparrow} \gamma_{\alpha -k \downarrow} |0\rangle$, where $|0\rangle$ is the vacuum for fermions. 
By the definition of the order parameter from Eq.~\eqref{equ:A1:eqd0}, the NH gap equation can be derived \cite{Suppl}. 
Notably, the gap equation retains the same form as in Eq.~\eqref{equ:A1:eqd255}, independent of $\phi$, demonstrating the gauge invariance of $\Delta_0$.
Explicitly,
\begin{equation}
\label{equ:A1:eqd18}
\Delta_0 = \frac{U}{4 N} \sum_{\alpha k} \frac{\Delta_0 + i(-1)^\alpha \left|w_k\right|}{\text{Re} E_{\alpha k} - i \gamma},
\end{equation}
which can be solved self-consistently by iterating the gap equation until convergence. In the Hermitian limit ($\gamma = 0$), Eq.~\eqref{equ:A1:eqd18} reduces to $\frac{1}{U} = \frac{1}{2N} \sum_k \frac{1}{E_k}$ where $E_k = \sqrt{|w_k|^2 + \Delta_0^2}$.

It is worth emphasizing that, as indicated by  Eq.~\eqref{equ:A1:eqd255}, $\Delta_0$ can be recognized as the order parameter when the $U(1)$ phase $\phi = 0$. In contrast to the Hermitian case, where $\Delta_0$ is purely real; here $\Delta_0 \in \mathbb{C}$, with both its real and imaginary parts being gauge-invariant. This distinction plays a crucial role in determining the properties of the NH superconductor.

\mytitle{First-order superconducting $\mathcal{PT}$ phase transition}.--By numerically solving the NH gap equation [Eq.~\eqref{equ:A1:eqd18}], we obtain the dependence of $\Delta_0$, $|\Delta_0|$ and $E_{\text{gap}}$ on $\gamma$ at a fixed $U$, as shown in Fig.~\ref{fig:fig2}(a) and (b). 
The spectrum gap $E_{\text{gap}}$ represents the energy required to break a Cooper pair, reflecting the ease of exciting a Bogoliubov quasiparticle.
In the $\mathcal{PT}$-symmetric phase, the spectrum gap is well defined and $E_{\text{gap}}= \sqrt{(\text{Re} \Delta_0)^2 - \gamma^2}$. 
We find that the gain and loss progressively reduce the spectrum gap, which eventually closes as the system transitions into the $\mathcal{PT}$-broken phase. 
Here we note that unlike the Hermitian case, $\Delta_0 \neq E_{\text{gap}}$ in the current case. 
Even so, the distinct nonlinear dependence of $|\Delta_0|$ on $\gamma$ is closely tied to the gap-closing behavior, where the monotonic behavior of $|\Delta_0|$ changes near the gap-closing point.

Notably, as $\gamma$ increases, $\Delta_0$, $|\Delta_0|$ and $E_{\text{gap}}$ exhibits abrupt jumps at the same $\gamma$, as indicated by the gray circles in Fig.~\ref{fig:fig2}(a) and (b). Meanwhile, $E_{\text{gap}}$ undergoes a jump at the gap-closing point, indicating a $\mathcal{PT}$ symmetry transition at this critical point. 
To further investigate this, we calculated the quasiparticle excitation spectrum \cite{Suppl}. The spectrum is real before the jump, indicating a $\mathcal{PT}$-symmetric phase. 
After the jump, the spectrum becomes complex. 
Therefore, the jumps of these physical quantities consistently indicate a first-order phase transition, which is closely related to the $\mathcal{PT}$ symmetry. 
As is well known, superconducting phase transitions are typically second-order in nature, except in rare cases where finite-momentum pairing dominates, such as in the superconducting FFLO state \cite{PhysRev.135.A550,larkin1965inhomogeneous}.
However, the present case reveals a novel possibility where NH dissipation induces a superconducting phase transition which is first order in nature.

The dependence of $\text{Re} \Delta_0$ on $\gamma$ for various $U$ are shown in Fig.~\ref{fig:fig2}(c), indicating that the jump of $\text{Re}\Delta_0$ is universal. We provide strong numerical evidence for this discontinuity by analyzing the convergence of its magnitude $\delta \text{Re} \Delta_0$ with respect to an infinitesimal change in $\gamma$ near the critical point \cite{Suppl}. The magnitude of the convergence jump, $\delta \text{Re} \Delta_0$, for various values of $U$ is presented in Fig.~\ref{fig:fig2}(d). Notably, even in the weak coupling regime, close to the Hermitian critical interaction strength $U_c$ \cite{Uchoa2007}, $\delta \text{Re} \Delta_0$ remains finite, as illustrated in the inset of Fig.~\ref{fig:fig2}(d).
To elucidate the underlying mechanisms of this first-order phase transition, we analyze the NH gap equation in the context of the $\mathcal{PT}$ phase transition. By increasing $\gamma$, the system crosses an exceptional point, where the eigenenergies of certain momenta become purely imaginary. This transition drastically alters the mathematical structure of the gap equation (Eq.~\eqref{equ:A1:eqd18}) and ultimately causing the jump in $\Delta_0$.
This means that the first-order phase transition is intrinsically a superconducting $\mathcal{PT}$ phase transition.
Numerical evidence, as shown in Fig.~\ref{fig:fig2}(c) and Fig.~\ref{fig:fig1}(b), further confirms this.
In Fig.~\ref{fig:fig2}(c), the gray dashed line marking the $\mathcal{PT}$ phase transition, defined by $\text{Re} \Delta_0 = \gamma$, intersects exactly with the critical points where $\Delta_0$ jumps. 
In Fig.~\ref{fig:fig1}(b), the black crosses marking $\Delta_0$'s jumps coincide with the $\mathcal{PT}$ phase transition line (blue curve) determined by the spectral properties. 

\begin{figure}[t] 
\includegraphics[width=8.6cm]{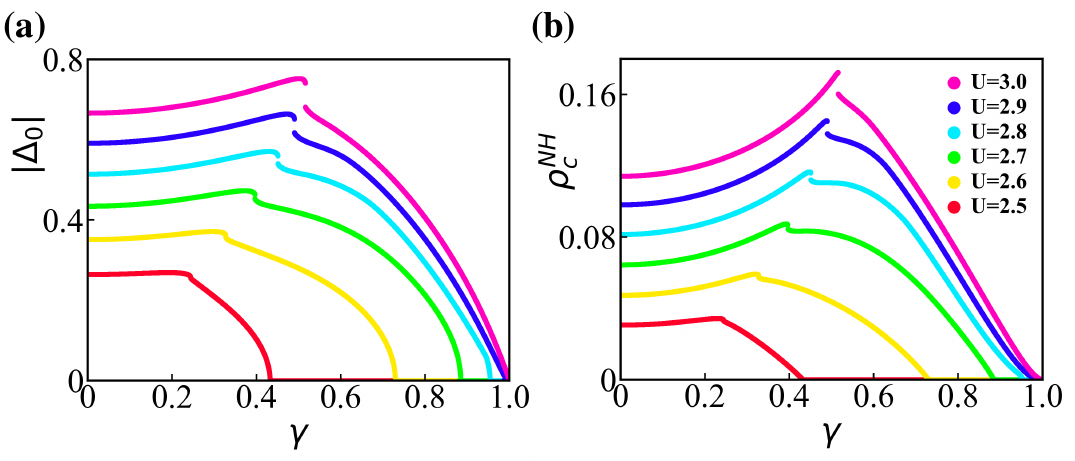} 
\caption{Experimentally measurable quantities, specifically (a) $|\Delta_0|$ and (b) $\rho_c^{NH}$, are plotted as functions of $\gamma$ for various values of $U$. Both quantities exhibit discontinuous jumps at identical critical parameters and display analogous responses to gain and loss. $|\Delta_0|$ can be obtained from localized spectral measurements, while $\rho_c^{NH}$ corresponds to macroscopic superfluid density measurements.}
\label{fig:fig3}
\end{figure}

\mytitle{Experimental observables}.--In the following discussion, we concentrate on the experimental observables that characterize the phase transition, beginning with localized spectral measurements. The random phase inherent in experimental data poses challenges for direct comparisons with $\Delta_0$, but these difficulties can be circumvented by analyzing its modulus, $|\Delta_0|$. Fig.~\ref{fig:fig3}(a) illustrates the dependence of $|\Delta_0|$ on $\gamma$ for different values of $U$.
As $\gamma$ increases, $|\Delta_0|$ exhibits discontinuous jumps and are accompanied by a drastic change in the monotonic behavior. 
For small $\gamma$, the system initially resides in the $\mathcal{PT}$-symmetric superconducting phase, $|\Delta_0|$
increases with $\gamma$, demonstrating that moderate dissipation can even enhance the superconductivity, contrary to
the conventional view that dissipation will sabotage the
superconductivity. 
This phenomenon arises from dissipation modifies the band structure of the system and significantly enhance the density of states (DOS) at the Fermi
level, which is otherwise identically zero in the Hermitian limit \cite{Uchoa2007,PhysRevA.84.021806,PhysRevB.90.094516}. As $\gamma$ continues to increase, the system transitions into the $\mathcal{PT}$-broken superconducting phase, where the spectrum gap closes, rendering the system highly susceptible to NH perturbations \cite{Suppl}. Consequently, the advantage of the enhanced DOS is ultimately outweighed by the detrimental impact of dissipation, leading to a rapid depletion of $|\Delta_0|$ until enters the normal phase with $|\Delta_0| = 0$.

The superconducting state can also be characterized by the existence of off-diagonal long-range order (ODLRO) in the second-order reduced density matrix $\rho_2(\vec{R}, \vec{R}^{\prime}, \vec{r})=\left\langle\psi_{\uparrow}^{\dagger}(\vec{R}^{\prime}+\vec{r}) \psi_{\downarrow}^{\dagger}(\vec{R}^{\prime}) \psi_{\downarrow}(\vec{R}) \psi_{\uparrow}(\vec{R}+\vec{r})\right\rangle$.
Here, $\vec{R}$ and $\vec{R}^{\prime}$ represent the center-of-mass coordinates, while $\vec{r}$ is the relative coordinate of the particle pair \cite{penrose1956bose,Yang1962}. 
A system is said to exhibit ODLRO if $\rho_2(\vec{R}, \vec{R}^{\prime}, \vec{r}) \not\to 0$ as $\left|\vec{R}-\vec{R}^{\prime}\right| \rightarrow \infty$. The spatial integral of $\rho_2(\vec{R}, \vec{R}^{\prime}, \vec{r})$ quantifies the density of condensed pairs, $\rho_c$, which is directly related to the macroscopic superfluid density measurements \cite{Suppl,pitaevskii2016bose}. In the $\mathcal{PT}$NHS model under investigation, $\rho_c^{NH}$ is explicitly given by
\begin{equation}
\label{equ:A1:eqdNc}
\rho_c^{NH}=\frac{1}{8 N} \sum_k\left|\sum_\alpha \frac{\Delta_0+i(-1)^\alpha\left|w_k\right|}{\operatorname{Re} E_{\alpha k}-i r}\right|^2.
\end{equation}
The dependence of $\rho_c^{NH}$ on $\gamma$ for various values of $U$ is presented in Fig.~\ref{fig:fig3}(b). As $\gamma$ increases, $\rho_c^{NH}$ exhibits discontinuous jumps at the same critical parameter as $|\Delta_0|$, as shown in Fig.~\ref{fig:figs2}(c) in the supplemental materials \cite{Suppl}. Similar to the behavior of $|\Delta_0|$, these discontinuous jumps in $\rho_c^{NH}$ are accompanied by a pronounced reversal in its monotonic trend. Consequently, the first-order superconducting $\mathcal{PT}$ phase transition can also be observed through macroscopic superfluid density measurements. The predicted abrupt jump in $\rho_c^{NH}$, corresponds to the superfluid stiffness, which can be probed via time-of-flight measurements in cold atoms, and in solid-state realizations through magnetic penetration depth measurements.

The $\mathcal{PT}$ phase diagram is presented in Fig.~\ref{fig:fig1}(b). As $\gamma$ increases for a fixed $U$, the system undergoes a general $\mathcal{PT}$ phase transition, indicated by the blue line in Fig.~\ref{fig:fig1}(b), which can be identified through indicators such as $|\Delta_0|$, $\rho_c^{NH}$, $\text{Re}\Delta_0$, and $E_{\text{gap}}$ \cite{Suppl}. 
Importantly, in the absence of pairing interactions, the $\mathcal{PT}$ symmetry would break under infinitesimal gain and loss. Superconducting pairing interactions protects the $\mathcal{PT}$ symmetry by opening a gap, with stronger pairing requiring larger critical values of $\gamma$ for the $\mathcal{PT}$ phase transition.
The competition between NH dissipation and superconducting pairing interactions ultimately governs the $\mathcal{PT}$ phase transition, as schematically illustrated in Fig.~\ref{fig:fig1}(a).
In the $\mathcal{PT}$-symmetric superconducting phase, moderate gain and loss enhance  superconductivity. 
In contrast, in the $\mathcal{PT}$-broken superconducting phase, Cooper pairs persist, but the spectrum gap closes, leaving the Cooper pairs highly vulnerable to NH dissipation and ultimately vanishing.
The red curve in Fig.~\ref{fig:fig1}(b) denotes the phase transition to the normal phase. 
In the strong dissipation limit, the quasiparticle eigenenergies $E_{\alpha k}$ become purely imaginary, making the gap equation have only a trivial solution.
Regardless of the strength of the pairing interaction, the system cannot form Cooper pairs, remaining in the normal phase.

\begin{figure}[t] 
\includegraphics[width=8.6cm]{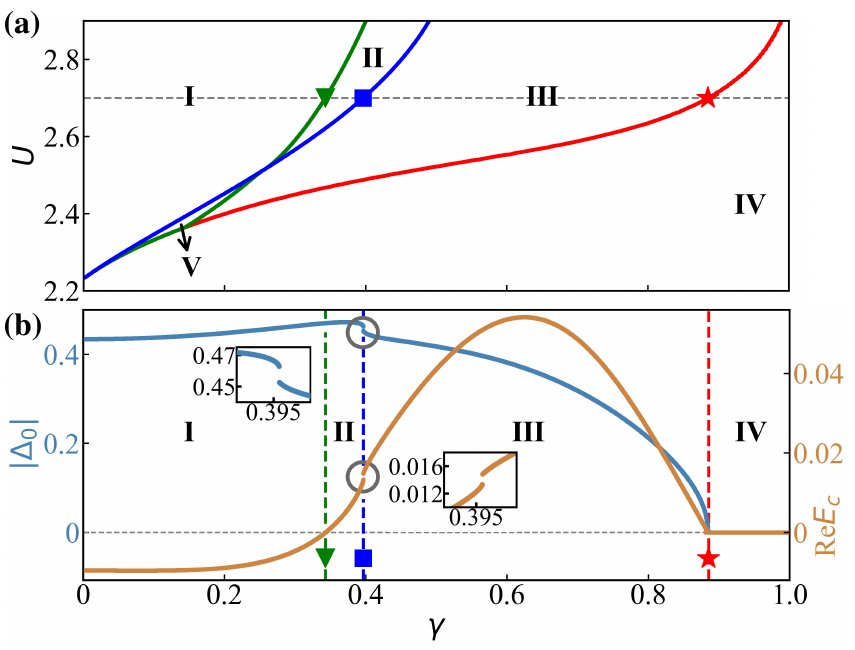}
\caption{(a) Phase diagram of the $\mathcal{PT}$NHS system at half-filling. The green, blue, and red curves denote the critical lines of the stability, $\mathcal{PT}$ symmetry, and superconducting to normal phase transitions, respectively. The identified phases are as follows: Phase I: $\mathcal{PT}$-symmetric stable superconducting phase; Phase II: $\mathcal{PT}$-symmetric metastable superconducting phase; Phase III: $\mathcal{PT}$-broken metastable superconducting phase; Phase IV: $\mathcal{PT}$-broken normal phase; Phase V: $\mathcal{PT}$-broken stable superconducting phase. (b) $\text{Re}\Delta_0$ (skyblue) and $\text{Re}E_c$ (brown) as functions of $\gamma$ for $U=2.7$, illustrating the phase transition along the gray dashed line in Fig.~\ref{fig:fig4}(a). Discontinuous jumps are marked by gray circles, with enlarged views provided in the insets.}
\label{fig:fig4}
\end{figure}

\mytitle{Stability of superconductivity}.--To evaluate the stability of the superconducting state as a reference for experimental observations of the superconducting $\mathcal{PT}$ phase transition, we compared the ground-state energies of the superconducting and normal states. Their difference is defined as the condensation energy:
\begin{equation}
E_c=\frac{2 }{U}|\Delta_0|^2-\frac{1 }{N}\sum_{\alpha k}E_{\alpha k}+\frac{2 }{N} \sum_{k} \sqrt{\left|w_k\right|^2-\gamma^2}.
\end{equation}
We note that gain and loss can destabilize the superconducting state, leading to a metastable superconducting state with $\text{Re} E_c > 0$ \cite{Suppl, Yamamoto2019}.

The phase diagram, which incorporates both the stability of superconductivity and $\mathcal{PT}$ symmetry, is presented in Fig.~\ref{fig:fig4}(a). The green curve represents the stability transition: the region to its left corresponds to the stable superconducting phase with $\text{Re} E_c < 0$, while the region to its right indicates the metastable superconducting phase with $\text{Re} E_c > 0$. The blue curve denotes the $\mathcal{PT}$ phase transition, and the red line marks the transition from the superconducting phase to the normal phase.
Fig.~\ref{fig:fig4}(b) provides a detailed depiction of the phase transitions along the gray dashed line in Fig.~\ref{fig:fig4}(a). Five distinct phases are identified, with stable superconducting phases, such as Phase I and Phase V, serving as promising platforms for experimental observation of the superconducting $\mathcal{PT}$ phase transition.

\mytitle{Summary.}--Our study reveals a novel first-order phase transition intrinsic to $\mathcal{PT}$NHS systems: the transition between the $\mathcal{PT}$-symmetric superconducting phase and the $\mathcal{PT}$-broken superconducting phase. This transition arises from the intricate interplay between superconducting pairing interactions and non-Hermitian dissipation. In the $\mathcal{PT}$-symmetric superconducting phase, moderate gain and loss enhance  superconductivity. Conversely, in the $\mathcal{PT}$-broken superconducting phase, strong dissipation significantly suppresses superconductivity. These phenomena can be experimentally probed through local spectral measurements and macroscopic superfluid density. The abrupt first-order nature of the transition near the $\mathcal{PT}$ breaking point suggests potential applications in quantum sensing, possibly opening new avenues for utilizing interacting non-Hermitian systems.
 
\let\oldaddcontentsline\addcontentsline
\renewcommand{\addcontentsline}[3]{}

\mytitle{Acknowledgements.}--This work was financially supported by the National Natural Science Foundation of China (Grants No.~12374037 and No.~12204044), and the Strategic Priority Research Program of the Chinese Academy of Sciences (Grant No. XDB28000000), and the Fundamental Research Funds for the Central Universities.

\bibliography{reference.bib}  
\let\addcontentsline\oldaddcontentsline

\newpage
\onecolumngrid
\newpage
{
	\center \bf \large 
	Supplemental Materials\\
	\large for ``\newtitle"\vspace*{0.1cm}\\ 
	\vspace*{0.5cm}
}

\tableofcontents

\setcounter{equation}{0}
\setcounter{figure}{0}
\setcounter{table}{0}
\setcounter{page}{1}

\renewcommand{\theequation}{S\arabic{equation}}
\renewcommand{\thefigure}{S\arabic{figure}}
\renewcommand{\theHtable}{Supplement.\thetable}
\renewcommand{\theHfigure}{Supplement.\thefigure}
\renewcommand{\bibnumfmt}[1]{[S#1]}

\section{The right-eigenstate-based non-Hermitian mean-field theory}

\begin{figure}[hbt] 
\includegraphics[width=12.6cm]{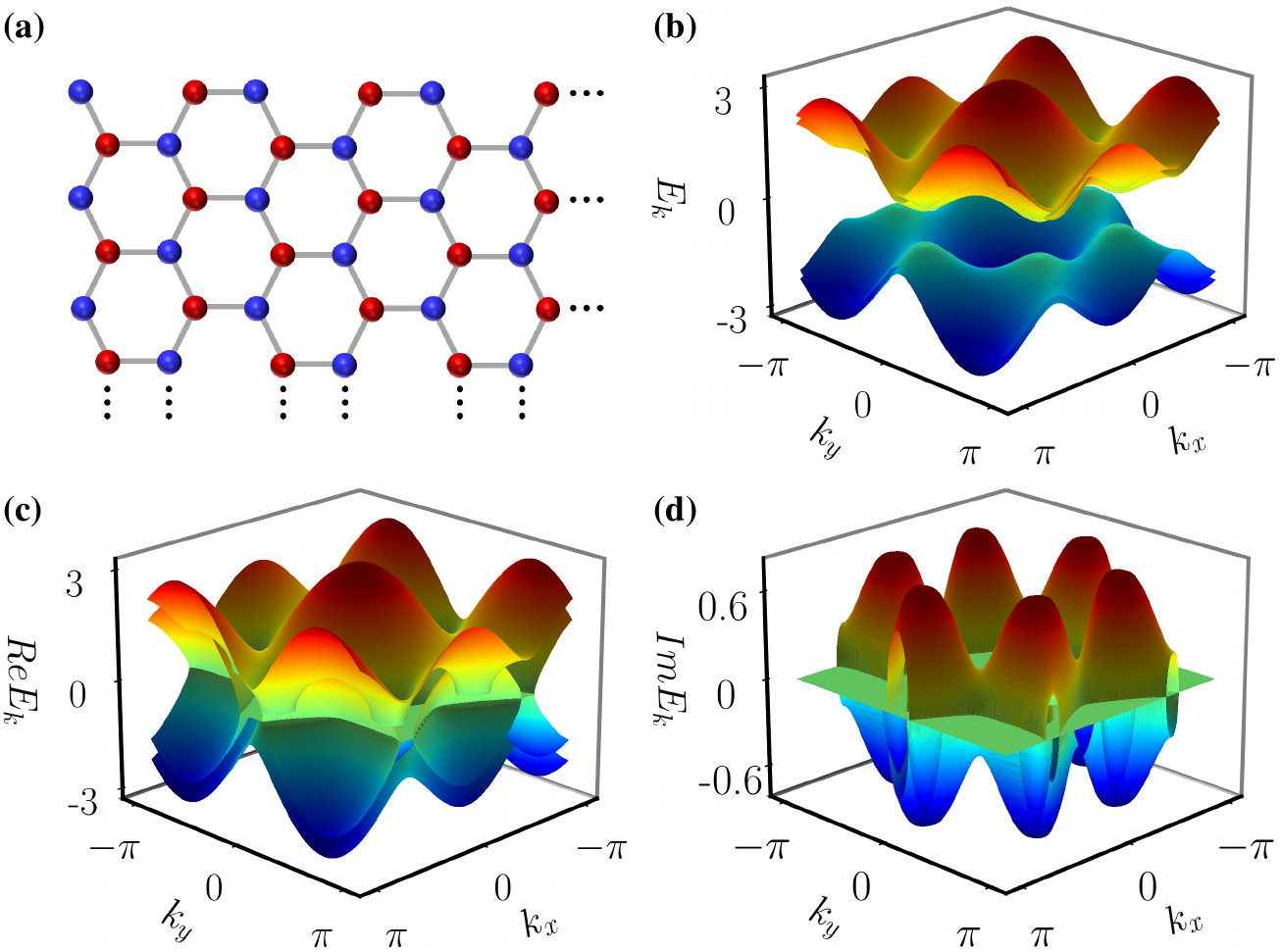}
\caption{(a) Schematic representation of the system under consideration. The red and blue balls represent gain and loss, respectively. (b) Quasiparticle excitation spectrum in the $\mathcal{PT}$-symmetric regime with $U=2.8$ and $\gamma=0.2$. (c) Real part and (d) imaginary part of the quasiparticle excitation spectrum in the $\mathcal{PT}$-broken regime with $U=2.8$ and $\gamma=0.8$.}
\label{fig:figs1}
\end{figure}

We focus on a $\mathcal{PT}$-symmetric non-Hermitian superconducting ($\mathcal{PT}$NHS) system described by
\begin{equation}
\label{equ:A1:eqsH}
\begin{aligned}
H= & -w \sum_{\langle n, m\rangle \sigma} c_{n \sigma}^{\dagger} c_{m \sigma}-(\mu+ i s_n \gamma) \sum_{n \sigma} c_{n \sigma}^{\dagger} c_{n \sigma} -U \sum_n c_{n \uparrow}^{\dagger} c_{n \downarrow}^{\dagger} c_{n \downarrow} c_{n \uparrow},
\end{aligned}
\end{equation}
where the non-Hermiticity arises from the alternating gain and loss on the $a\ (s_n=1)$ and $b\ (s_n=-1)$ sublattices, denoted by $\gamma$, as illustrated in Fig.~\ref{fig:figs1}(a). For simplicity, we consider the half-filling case, where the chemical potential $\mu$ at the Fermi surface is zero. 

The Lindblad quantum master equation describes the evolution of a system coupled to its environment
\begin{equation}
\label{equ:A1:eqsH223}
\begin{aligned}
\frac{d\rho_s(t)}{dt} & =-i[H_0, \rho_s(t)]-\frac{1}{2}  \sum_n \gamma_n\left  [ L_n^{\dagger} L_n \rho_s(t)+\rho_s(t) L_n^{\dagger} L_n-2 L_n \rho_s(t) L_n^{\dagger} \right ] ,
\end{aligned}
\end{equation}
where $\rho_s$ is the reduced density matrix and $L_n$ is the Lindblad operator representing a loss at site $n$ with strength $\gamma_n$ \cite{PhysRevLett.68.580}. When the dynamical evolution timescale satisfies $t \ll 1/\gamma$, the quantum jump effects described by the last term in Eq.~\eqref{equ:A1:eqsH223} can be neglected. Under this condition, the system can be effectively described by the NH Hamiltonian
\begin{equation}
H_{\mathrm{eff}}=H_0-\frac{i}{2}   \sum_n \gamma_n L_n^{\dagger}  L_n.
\end{equation}
Thus, Eq.~\eqref{equ:A1:eqsH} describes the one-body loss processes induced by coupling to the environment, where the Lindblad operator is defined as $L_n = c_n$ and the dissipation strength $\gamma_n$ are different for the two sublattices.

In the NH quantum mechanics, the orthogonality of eigenstates are broken. In addition, the evolution of the state is nonunitary
\begin{equation}
\langle\psi(t) \mid \psi(t)\rangle=\langle\psi(0)| U^{\dagger}(t) U(t)|\psi(0)\rangle=\langle\psi(0)| e^{i\left(H^{\dagger}-H\right) t}|\psi(0)\rangle \ne 1.
\end{equation}
Therefore, the expectation value of operator $O$ at the state $|\psi(t)\rangle$ is given by 
\begin{equation}
\label{equ:A1:eqd10}
\langle O\rangle=\frac{\langle\psi(t)| O|\psi(t)\rangle}{\langle\psi(t) \mid \psi(t)\rangle} ,
\end{equation}
where the denominator is proper normalization, maintaining the conservation of probability\cite{Sticlet2022}. This definition follows the ideas from the Hermitian quantum mechanics and is both experimentally more relevant \cite{Barontini2013,Li2019,Wu2019} and physically meaningful \cite{Meden2023,Sticlet2022,Yan2024,Dora2019,Yamamoto2021,Yamamoto2022}. We refer to Eq.~\eqref{equ:A1:eqd10} as the right-eigenstate-based definition. Additionally, there exists another definition based on the biorthogonal basis \cite{Brody2013,Yamamoto2019}. While this biorthogonal definition offers a mathematically concise form for defining observables, its physical significance and experimental observability remain ambiguous. 
Studies on normal metal-insulator-$ \mathcal{PT}$NHS junction have shown that, compared to the biorthogonal definition, the right-eigenstate-based definition more accurately captures the Andreev reflection process \cite{Kornich2023}.

Therefore, we develope a right-eigenstate-based NH mean-field theory to describe the properties of NH superconductors. The definition of the order parameters are as follows:
\begin{equation}
\begin{aligned}
\label{equ:A1:eqs11}
& \Delta_a=\frac{U}{N} \sum_k \left\langle a_{-k \downarrow} a_{k \uparrow}\right\rangle \equiv \frac{U}{N} \sum_k \frac{\left\langle \Psi_0\left|a_{-k \downarrow} a_{k \uparrow}\right| \Psi_0\right\rangle}{\langle \Psi_0 \mid \Psi_0\rangle}, \\[2mm]
& \Delta_b=\frac{U}{N} \sum_k \left\langle b_{-k \downarrow} b_{k \uparrow}\right\rangle \equiv\frac{U}{N} \sum_k \frac{\left\langle \Psi_0\left|b_{-k \downarrow} b_{k \uparrow}\right| \Psi_0\right\rangle}{\langle \Psi_0 \mid \Psi_0\rangle}, \\[2mm]
\end{aligned}
\end{equation}
where $|\Psi_0\rangle$ is the NH BCS ground state.
After perform the right-eigenstate-based NH mean-field approximation, we obtain
\begin{equation}
\label{equ:A1:eqsHmf}
\begin{aligned}
H_{MF} = & -\sum_{k \sigma}\left(w_k a_{k \sigma}^{\dagger} b_{k \sigma} + \text{h.c.} \right) - i \gamma \sum_{k \sigma}\left( a_{k \sigma}^{\dagger} a_{k \sigma} - b_{k \sigma}^{\dagger} b_{k \sigma} \right)\\
 &-\sum_k\left[\left(\Delta_a a_{k \uparrow}^{\dagger} a_{-k \downarrow}^{\dagger}+\Delta_b b_{k \uparrow}^{\dagger} b_{-k \downarrow}^{\dagger}\right)+\text{h.c.}\right]+ \frac{2 N}{U}|\Delta_0|^2,
\end{aligned}
\end{equation}
where $w_k=w\left[e^{-i k_y}+e^{i(\frac{\sqrt{3}}{2}  k_x + \frac{1}{2} k_y)}+e^{i(-\frac{\sqrt{3}}{2}  k_x+\frac{1}{2} k_y)}\right]$. 
Although $\Delta_a \ne \Delta_b$, their gauge phases are identical according to Eq.~\eqref{equ:A1:eqs11}.
Thus, considering the alternating gain and loss on the sublattices and the U(1) symmetry of the system's Hamiltonian [Eq.~\eqref{equ:A1:eqsH}], we can generally define
\begin{equation}
\label{equ:A1:eqs255}
\begin{aligned}
\Delta_a=\Delta_0 e^{i \phi}, \ \ \Delta_b=\Delta_0^* e^{i \phi},
\end{aligned}
\end{equation}
where $\Delta_0 \in \mathbb{C}$, $\Delta_a$ and $\Delta_b$ remain independent of each other. 

Due to the non-Hermiticity, the NH mean-field Hamiltonian [Eq.~\eqref{equ:A1:eqsHmf}] cannot be diagonalized by a unitary transformation. To diagonalize it, we define the NH Bogoliubov transformation which is given by
\begin{equation}
\label{equ:A1:eqs12}
\begin{aligned}
& \bar{\gamma}_{1 k \uparrow}=\frac{1}{\sqrt{2}}\left(e^{i \phi} \cdot i e^{i \theta_k} v_{1 k} a_{k \uparrow}^{\dagger}-e^{i \phi} \cdot i \bar{v}_{1 k}^\prime b_{k \uparrow}^{\dagger}+e^{i \theta_k} \bar{v}_{1 k}^\prime a_{-k \downarrow}+v_{1 k} b_{-k \downarrow}\right), \\
& \bar{\gamma}_{2 k \uparrow}=\frac{1}{\sqrt{2}}\left(-e^{i \phi} \cdot i e^{i \theta_k} v_{2 k} a_{k \uparrow}^{\dagger}+e^{i \phi} \cdot i \bar{v}_{2 k}^\prime b_{k \uparrow}^{\dagger}+e^{i \theta_k} \bar{v}_{2 k}^\prime a_{-k \downarrow}+v_{2 k} b_{-k \downarrow}\right) ,\\
& \bar{\gamma}_{1-k \downarrow}=\frac{1}{\sqrt{2}}\left(-e^{-i \phi} \cdot i e^{-i \theta_k} u_{1 k} a_{k \uparrow}-e^{-i \phi} \cdot i \bar{v}_{1 k} b_{k \uparrow}-e^{-i \theta_k} \bar{v}_{1 k} a_{-k \downarrow}^{\dagger}+u_{1 k} b_{-k \downarrow}^{\dagger}\right), \\
& \bar{\gamma}_{2-k \downarrow}=\frac{1}{\sqrt{2}}\left(e^{-i \phi} \cdot i e^{-i \theta_k} u_{2 k} a_{k \uparrow}+e^{-i \phi} \cdot i \bar{v}_{2 k} b_{k \uparrow}-e^{-i \theta_k} \bar{v}_{2 k} a_{-k \downarrow}^{\dagger}+u_{2 k} b_{-k \downarrow}^{\dagger}\right),\\
& \gamma_{1 k \uparrow}=\frac{1}{\sqrt{2}}\left(-e^{-i \phi} \cdot i e^{-i \theta_k} v_{1 k} a_{k \uparrow}+e^{-i \phi} \cdot i \bar{u}_{1 k} b_{k \uparrow}+e^{-i \theta_k} \bar{u}_{1 k} a_{-k \downarrow}^{\dagger}+v_{1 k} b_{-k \downarrow}^{\dagger}\right), \\
& \gamma_{2 k \uparrow}=\frac{1}{\sqrt{2}}\left(e^{-i \phi} \cdot i e^{-i \theta_k} v_{2 k} a_{k \uparrow}-e^{-i \phi} \cdot i \bar{u}_{2 k} b_{k \uparrow}+e^{-i \theta_k} \bar{u}_{2 k} a_{-k \downarrow}^{\dagger}+v_{2 k} b_{-k \downarrow}^{\dagger}\right), \\
& \gamma_{1-k \downarrow}=\frac{1}{\sqrt{2}}\left(e^{i \phi} \cdot i e^{i \theta_k} u_{1 k} a_{k \uparrow}^{\dagger}+e^{i \phi} \cdot i \bar{u}_{1 k}^\prime b_{k \uparrow}^{\dagger}-e^{i \theta_k} \bar{u}_{1 k}^\prime a_{-k \downarrow}+u_{1 k} b_{-k \downarrow}\right) ,\\
& \gamma_{2-k \downarrow}=\frac{1}{\sqrt{2}}\left(-e^{i \phi} \cdot i e^{i \theta_k} u_{2 k} a_{k \uparrow}^{\dagger}-e^{i \phi} \cdot i \bar{u}_{2 k}^\prime b_{k \uparrow}^{\dagger}-e^{i \theta_k} \bar{u}_{2 k}^\prime a_{-k \downarrow}+u_{2 k} b_{-k \downarrow}\right),
\end{aligned}
\end{equation}
where $\theta_k = Arg(w_k)$, and the transformation coefficients are as follows:
\begin{equation}
\label{equ:A1:eq15}
\begin{aligned}
 u_{\alpha k}=\sqrt{\frac{E_{\alpha k}+i \gamma}{2 E_{\alpha k}}},&\ v_{\alpha k}=\sqrt{\frac{E_{\alpha k}-i \gamma}{2 E_{\alpha k}}},\\[2mm]
\ \bar{u}_{\alpha k}=\frac{E_{\alpha k}+i \gamma}{\left|w_k\right|+i(-1)^\alpha \Delta_0^*} \sqrt{\frac{E_{\alpha k}-i \gamma}{2 E_{\alpha k}}},& \ \bar{v}_{\alpha k}=\frac{E_{\alpha k}-i \gamma}{\left|w_k\right|+i(-1)^\alpha \Delta_0^*} \sqrt{\frac{E_{\alpha k}+i \gamma}{2 E_{\alpha k}}},\\[2mm]
 \ \bar{u}_{\alpha k}^\prime=\frac{E_{\alpha k}-i \gamma}{\left|w_k\right|-i(-1)^\alpha \Delta_0} \sqrt{\frac{E_{\alpha k}+i \gamma}{2 E_{\alpha k}}},& \ \bar{v}_{\alpha k}^\prime=\frac{E_{\alpha k}+i \gamma}{\left|w_k\right|-i(-1)^\alpha \Delta_0} \sqrt{\frac{E_{\alpha k}-i \gamma}{2 E_{\alpha k}}}, 
\end{aligned}
\end{equation}
where $\alpha=1,2$ labels the branches of the quasiparticles, $E_{\alpha k}$ is the excitation energy of them.
What's fascinating is that even though the definition of the NH Bogoliubov transformation is divided into four cases depending on the $\mathcal{PT}$ symmetry, they all have the same mathematical form ultimately.

After perform the NH Bogoliubov transformation, Eq.~\eqref{equ:A1:eqsHmf} can be diagonalized as
\begin{equation}
\label{equ:A1:eq7}
\begin{aligned}
\hat{H}_{MF} & =\sum_k\left(\begin{array}{llll}
\bar{\gamma}_{1 k \uparrow} & \bar{\gamma}_{2 k \uparrow} & \gamma_{1-k \downarrow} & \gamma_{2-k \downarrow}
\end{array}\right)\left(\begin{array}{cccc}
E_{1 k} & 0 & 0 & 0 \\
0 & E_{2 k} & 0 & 0 \\
0 & 0 & -E_{1 k} & 0 \\
0 & 0 & 0 & -E_{2 k}
\end{array}\right)\left(\begin{array}{c}
\gamma_{1 k \uparrow} \\
\gamma_{2 k \uparrow} \\
\bar{\gamma}_{1-k \downarrow} \\
\bar{\gamma}_{2-k \downarrow}
\end{array}\right) + \frac{2 N}{U}|\Delta_0|^2,
\end{aligned}
\end{equation}
where $\bar \gamma_{\alpha k\sigma}$ and $\gamma_{\alpha k\sigma}$ are the creation and annihilation operator of quasiparticles, and $E_{1k}= \sqrt {{{\left| {\left| w_k \right| + i\Delta_0 } \right|}^2} - \gamma ^2}$ and $E_{2k} = \sqrt {{{\left| {\left| w_k \right| - i\Delta_0 } \right|}^2} - \gamma ^2}$ are the excitation energy of quasiparticles. Therefore, \(\text{Re} \Delta_0 \ge \gamma\) ensures that \(\mathcal{PT}\) symmetry is preserved, while \(\text{Re} \Delta_0 < \gamma\) indicates that \(\mathcal{PT}\) symmetry is broken. The quasiparticle excitation spectrum  for both the $\mathcal{PT}$-symmetric and $\mathcal{PT}$-broken regimes are presented in Fig.~\ref{fig:figs1} (b), (c), and (d). Different from the Hermitian case, $\bar\gamma_{\alpha k\sigma} \ne \gamma_{\alpha k\sigma}^{\dagger} $ in current case. Therefore, although satisfy $\left \{\bar\gamma_{\alpha k\sigma},\gamma_{\alpha'k'\sigma'}\right \}   = \delta_{\alpha \alpha'}\delta_{kk'}\delta_{\sigma\sigma'}$, these quasiparticles are neither conventional fermions nor bosons.

In the ground state, all quasiparticle excitations are absent.
Consequently, $|\Psi_0\rangle = \prod_{\alpha,k} \gamma_{\alpha k \uparrow} \gamma_{\alpha -k \downarrow} |0\rangle$, where $|0\rangle$ is the vacuum for fermions.
With Eq.~\eqref{equ:A1:eqs12}, the ground state $|\Psi_0\rangle $ can be rewritten as
\begin{equation}
\label{equ:A1:eq19}
\begin{array}{l}
|\Psi_0\rangle  = \prod\limits_k {\left[ {\frac{1}{2}\left( {v_{1 k}\bar u_{2 k} - \bar u_{1 k}v_{2 k}} \right)|{\rm{0}}\rangle }   - \frac{1}{2} e^{i 2\phi}\cdot\left( {v_{1 k}\bar u_{2 k} - \bar u_{1 k}v_{2 k}} \right)a_{ - k \downarrow }^\dagger b_{ - k \downarrow }^\dagger a_{k \uparrow }^\dagger b_{k \uparrow }^\dagger |{\rm{0}}\rangle \right.}\\[4mm]
\qquad \qquad  + \frac{1}{2}e^{i \phi}\cdot\left( {v_{1 k}\bar u_{2 k} + \bar u_{1 k}v_{2 k}} \right)\left( {i{e^{ - i\theta_k }}a_{ - k \downarrow }^\dagger b_{k \uparrow }^\dagger |{\rm{0}}\rangle  + i{e^{i\theta_k }}b_{ - k \downarrow }^\dagger a_{k \uparrow }^\dagger |{\rm{0}}\rangle } \right)\\[4mm]
\qquad \qquad \  \left.{ + ie^{i \phi}\cdot\left( {\bar u_{1 k}\bar u_{2 k}a_{ - k \downarrow }^\dagger a_{k \uparrow }^\dagger |{\rm{0}}\rangle  + v_{1 k}v_{2 k}b_{ - k \downarrow }^\dagger b_{k \uparrow }^\dagger |{\rm{0}}\rangle } \right)} \right],
\end{array}
\end{equation}
with the normalization coefficient $\langle \Psi_0 \mid \Psi_0\rangle=\prod\limits_k \left(\left|u_{1 k}\right|^2+\left|v_{1 k}\right|^2\right) \left(\left|u_{2 k}\right|^2+\left|v_{2 k}\right|^2\right)$. In the Hermitian limit, we have $\left|u_{\alpha k}\right|^2+\left|v_{\alpha k}\right|^2 =1$.
Substituting Eq.~\eqref{equ:A1:eq19} into Eq.~\eqref{equ:A1:eqs11}, we can derive the NH gap equation
\begin{equation}
\label{equ:A1:eqsge}
\begin{aligned}
& \Delta_a=i \frac{U}{2 N} \sum_k\left(-\frac{\bar{u}_{1 k} v_{1 k}^*}{\left|u_{1 k}\right|^2+\left|v_{1 k}\right|^2}+\frac{\bar{u}_{2 k} v_{2 k}^*}{\left|u_{2 k}\right|^2+\left|v_{2 k}\right|^2}\right) \cdot e^{i \phi}, \\[2mm]
& \Delta_b=-i \frac{U}{2 N} \sum_k\left(-\frac{\bar{u}_{1 k}^* v_{1 k}}{\left|u_{1 k}\right|^2+\left|v_{1 k}\right|^2}+\frac{\bar{u}_{2 k}^* v_{2 k}}{\left|u_{2 k}\right|^2+\left|v_{2 k}\right|^2}\right) \cdot e^{i \phi},
\end{aligned}
\end{equation}
which still keep the form of Eq.~\eqref{equ:A1:eqs255}. 
Accordingly,
\begin{equation}
\label{equ:A1:eqsge0}
\begin{aligned}
& \Delta_0=i \frac{U}{2 N} \sum_k\left(-\frac{\bar{u}_{1 k} v_{1 k}^*}{\left|u_{1 k}\right|^2+\left|v_{1 k}\right|^2}+\frac{\bar{u}_{2 k} v_{2 k}^*}{\left|u_{2 k}\right|^2+\left|v_{2 k}\right|^2}\right).
\end{aligned}
\end{equation}
With Eq.~\eqref{equ:A1:eq15}, it can be rewritten as
\begin{equation}
\label{equ:A1:eqsgapeq}
\Delta_0= \frac{U}{4 N} \sum_k\left(\frac{ \Delta_0 - i\left|w_k\right|}{\text{Re} E_{1 k}-i r}+\frac{\Delta_0 + i\left|w_k\right|}{\text{Re} E_{2 k}-i r}\right),
\end{equation}
which is independent of $\phi$, demonstrating the gauge invariance of $\Delta_0$. 
$\Delta_0$ can be recognized as the order parameter when U(1) phase $\phi = 0$. 
Unlike $\Delta_a$ and $\Delta_b$, both the real and imaginary parts of $\Delta_0$ are gauge-invariants.
Thus, the first-order phase transition, characterized by a discontinuous jump of $\Delta_0$, is not a phenomenon tied to the gauge choice.

In the Hermitian limit, we have $\Delta_0  \in \mathbb{R}$ and $E_{\alpha k} = \sqrt{|\omega_k|^2 + \Delta_{0}^{2}}=E_k$. Then, the NH gap equation reduces to the Hermitian case
\begin{equation}
\Delta_0= \frac{U}{2 N} \sum_k \frac{\Delta_0}{E_k}.
\end{equation}

\section{The off-diagonal long-range order and the condensed density}

The presence of the off-diagonal long-range order (ODLRO) characterizes the macroscopic phase coherence, serving as the fundamental mechanism underlying superfluidity \cite{penrose1956bose,Yang1962}. 
For the second-order reduced density matrix $\rho_2(\vec{R}, \vec{R}^{\prime},  \vec{r})=\left\langle\psi_{\uparrow}^{\dagger}(\vec{R}^{\prime}+\vec{r}) \psi_{\downarrow}^{\dagger}(\vec{R}^{\prime}) \psi_{\downarrow}(\vec{R}) \psi_{\uparrow}(\vec{R}+\vec{r})\right\rangle$, if $\rho_2(\vec{R}, \vec{R}^{\prime}, \vec{r}) \not\to 0$ as $\left|\vec{R}-\vec{R}^{\prime}\right| \rightarrow \infty$, ODLRO is said to exist. 
Here, $\vec{R}$ and $\vec{R}^{\prime}$ represent the center-of-mass coordinates, while $\vec{r}$ is the relative coordinate of the particle pair. 

By assuming spontaneous gauge symmetry breaking and incorporating the translational symmetry intrinsic to the system, the pairing field can be introduced as follows:
\begin{equation}
\label{equ:A1:eqsff}
\begin{aligned}
F(\vec{r}) & =\left\langle\psi_{\downarrow}(\vec{R}) \psi_{\uparrow}(\vec{R}+\vec{r})\right\rangle \\
& =\frac{1}{N} \sum_{k}\left\langle c_{-k \downarrow} c_{k \uparrow}\right\rangle e^{i \vec{k} \cdot \vec{r}},
\end{aligned}
\end{equation}
where the dependence on the center-of-mass coordinates vanishes, enabling the second-order reduced density matrix to be rewritten as $\rho_2(\vec{R}, \vec{R}^{\prime}, \vec{r})=|F(\vec{r})|^2$. 
Consequently, the condensed density is defined as
\begin{equation}
\rho_c=\int d \vec{r} \ |F(\vec{r})|^2,
\end{equation}
which is directly related to the macroscopic superfluid density measurements \cite{pitaevskii2016bose}.

For the $\mathcal{PT}\mathrm{NHS}$ model under investigation, the condensed density $\rho_c^{NH}$ is explicitly expressed as $\rho_c^{N H}  =\frac{1}{N} \sum_{k}\left(\left\langle a_{k \uparrow}^{\dagger} a_{-k \downarrow}^{\dagger}\right\rangle\left\langle a_{-k \downarrow} a_{k \uparrow}\right\rangle+\left\langle b_{k \uparrow}^{\dagger} b_{-k \downarrow}^{\dagger}\right\rangle\left\langle b_{-k \downarrow} b_{k \uparrow}\right\rangle\right)$. By applying the definition of observables in Eq.~\eqref{equ:A1:eqd10} and incorporating the BCS ground state defined in Eq.~\eqref{equ:A1:eq19}, $\rho_c^{NH}$ can be further expressed as
\begin{equation}
\rho_c^{N H} =\frac{1}{8 N} \sum_k\left|\sum_\alpha \frac{\Delta_0+i(-1)^\alpha\left|w_k\right|}{\operatorname{Re} E_{\alpha k}-i r}\right|^2,
\end{equation}
where contributions from the $a$ and $b$ sublattices are equal.

\section{The existence and characterization of the first-order superconducting $\mathcal{PT}$ phase transition}

\begin{figure}[hbt]
\includegraphics[width=12.6cm]{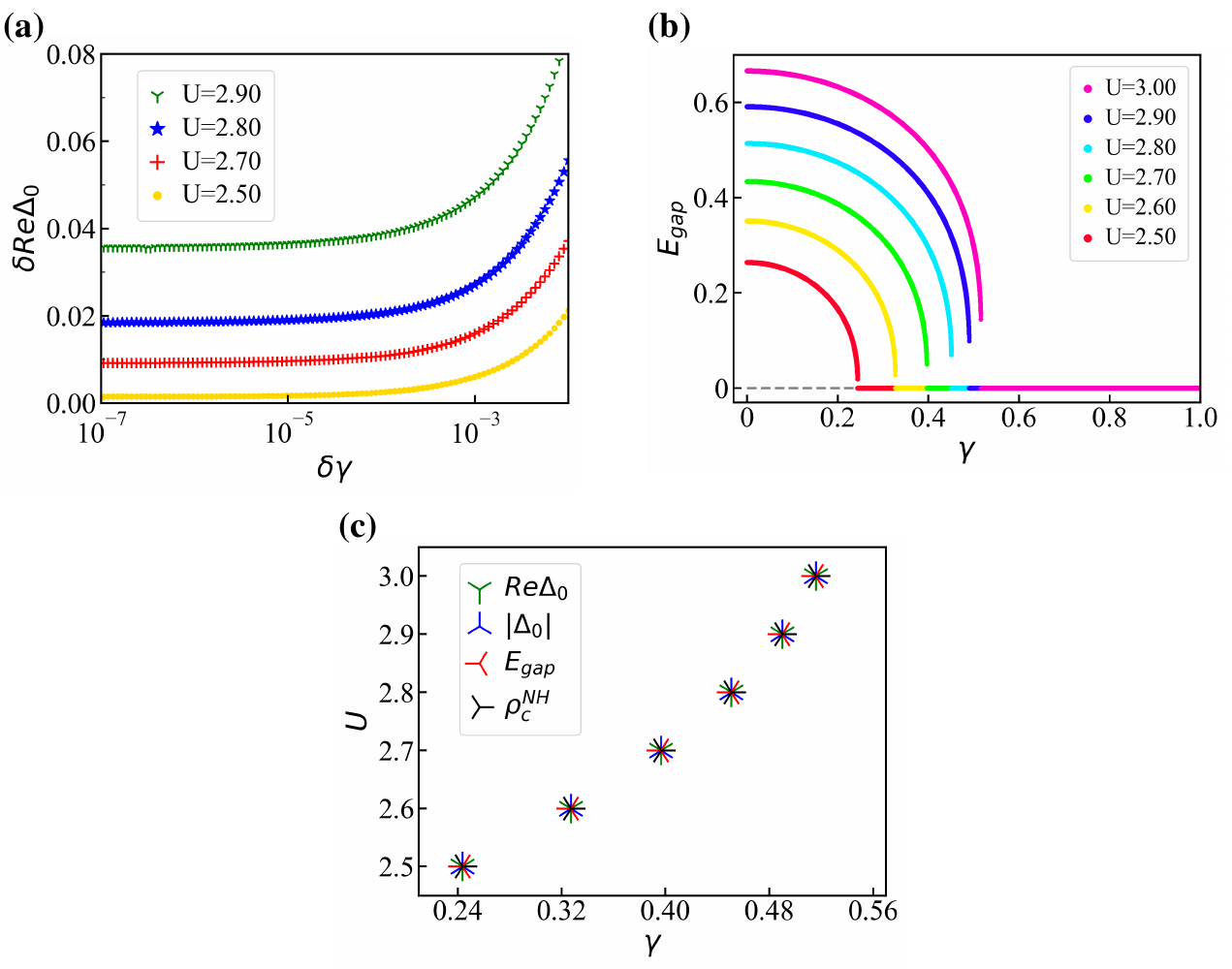}
\caption{(a) The jump magnitude $\delta \text{Re} \Delta_0$ as a function of $\delta\gamma$ around the critical point for various values of $U$. (b) $E_{\text{gap}}$ plotted against $\gamma$ for different $U$. (c) The critical $\gamma$ for different $U$ at the $\mathcal{PT}$ phase transition, determined by the jumps in $\text{Re}\Delta_0$, $|\Delta_0|$, $E_{\text{gap}}$ and $\rho_c^{NH}$, are shown to coincide.}
\label{fig:figs2}
\end{figure}

To verify the existence of the first-order phase transition, we calculate $\delta \text{Re} \Delta_0$ as a function of $\delta \gamma$ around the critical point for various pairing interaction strengths $U$, as shown in Fig.~\ref{fig:figs2} (a). 
As $\delta \gamma$ decreases, $\delta \text{Re} \Delta_0$ converged. 
The converged value is the jump magnitude, which increases monotonically with $U$.
This confirms that the discontinuous jump of $\Delta_0$ is not an artifact of numerical precision limitations. 
A series of jump magnitudes corresponding to different $U$ are presented in Fig.~\ref{fig:fig2}(d).

Moreover, we numerically calculate $E_{\text{gap}}$ as a function of $\gamma$ for various values of $U$, as shown in Fig.~\ref{fig:figs2} (b). 
As $\gamma$ increases, the spectrum gap $E_{\text{gap}}$ decreases rapidly and eventually closes following an abrupt jump, serving as a clear indicator of the first-order superconducting $\mathcal{PT}$ phase transition. 
This gap-closing behavior profoundly impacts the superconducting response to non-Hermitian dissipation, leading to a reversal in the monotonic behavior of $|\Delta_0|$ and $\rho_c^{NH}$. 
As illustrated in Fig.~\ref{fig:figs2} (c), the critical $\gamma$ for the $\mathcal{PT}$ phase transition, identified by the jumps in $\text{Re}\Delta_0$, $|\Delta_0|$, $E_{\text{gap}}$ and $\rho_c^{NH}$, aligns consistently across all indicators. These quantities collectively serve as robust and universal markers of the first-order superconducting $\mathcal{PT}$ phase transition, highlighting the generality and stability of the phase transition.

\section{Condensation energy}

\begin{figure}[hbt]
\includegraphics[width=12.6cm]{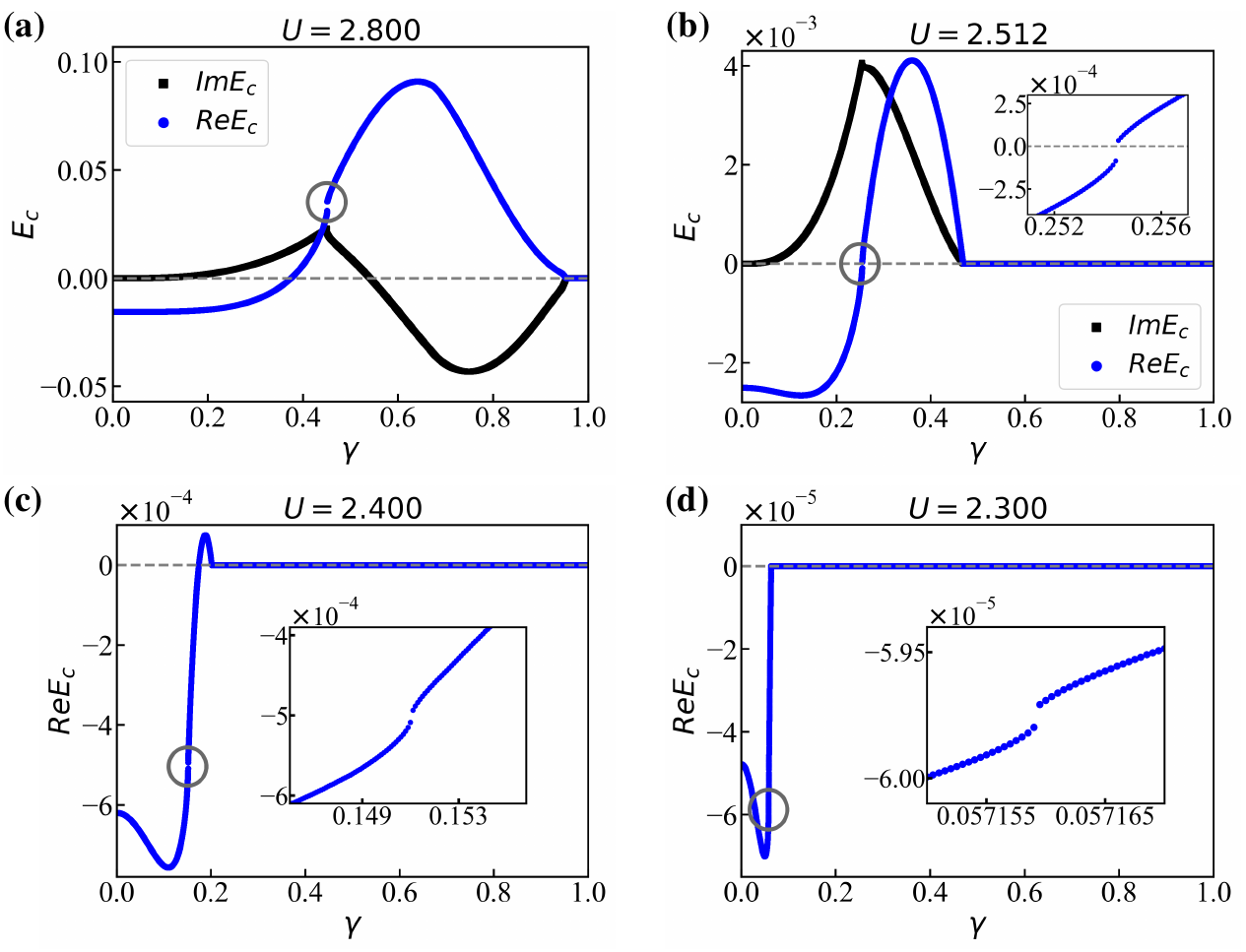}
\caption{Condensation energy $E_c$ as a function of $\gamma$ for different $U$. Gray circles indicate the discontinuous jumps, with insets for the enlarged views. The dashed lines represent $E_c=0$.}
\label{fig:figs4}
\end{figure}

In this section, we discuss the stability of the superconducting state through the analysis of the condensation energy. The condensation energy is defined as the difference between the ground-state energy of the superconducting and normal states.
For $\mu=0$, we can prove 
\begin{equation}
\begin{aligned}
& \sum_{k \sigma} \frac{\left\langle \Psi_0\left|a_{k \sigma}^{\dagger} a_{k \sigma}+b_{k \sigma}^{\dagger} b_{k \sigma}\right| \Psi_0\right\rangle}{\langle \Psi_0 \mid \Psi_0\rangle} \\[1mm]
=& \ 2 \sum_k  \frac{\left|u_{2 k}\right|^2\left|v_{1 k}\right|^2+\left|u_{1 k}\right|^2\left|v_{2 k}\right|^2+\left|u_{1 k}\right|^2\left|u_{2 k}\right|^2+\left|v_{1 k}\right|^2\left|v_{2 k}\right|^2}{\left(\left|u_{1 k}\right|^2+\left|v_{1 k}\right|^2\right)\left(\left|u_{2 k}\right|^2+\left|v_{2 k}\right|^2\right)} \\[1mm]
=& \ 2N,
\end{aligned}
\end{equation}
indicating that the system is at half-filling. Then, with Eq.~\eqref{equ:A1:eq7}, the condensation energy of single unit cell is given by
\begin{equation}
\label{equ:A1:eqs40}
E_c=\left [ \frac{2 N}{U}|\Delta_0|^2-\sum_{\alpha k}E_{\alpha k}+2 \sum_{k} \sqrt{\left|w_k\right|^2-\gamma^2} \right ] / N .
\end{equation}
Fig.~\ref{fig:figs4} shows the dependence of $E_c$ on $\gamma$ for different $U$.
Interestingly, under sufficiently strong dissipation, the superconducting state becomes metastable which shows the destablize effect on superconducting state from gain and loss, as shown in Fig.~\ref{fig:figs4}(a), (b), and (c).
In the weak dissipation region, gain and loss initially enhance the stability of the superconducting state, as shown in Fig.~\ref{fig:figs4}(b), (c), and (d).
This can be attributed to the fact that, unlike the superconducting state, the normal state lacks the protection of superconductivity, rendering it more vulnerable to the NH perturbation.

According to the definition provided in [Eq.~\eqref{equ:A1:eqs40}], there are also discontinuous jumps in the condensation energy at the critical points of the superconducting $\mathcal{PT}$ phase transition, as indicated by the gray circles in Fig.~\ref{fig:figs4}.
Thus, the real part of the condensation energy can characterize both stability of superconductivity and $\mathcal{PT}$ symmetry to some extent. 
Fig.~\ref{fig:figs4} shows the phase transitions regulated by gain and loss under different $U$, corresponding to the phase diagram in Fig.~\ref{fig:fig4}(a). 
Specifically, Fig.~\ref{fig:figs4}(a) undergoes: Phase I - Phase II - Phase III - Phase IV; Fig.~\ref{fig:figs4}(b) undergoes:  Phase I - Phase III - Phase IV; Fig.~\ref{fig:figs4}(c) undergoes:  Phase I - Phase V - Phase III - Phase IV; Fig.~\ref{fig:figs4}(d) undergoes:  Phase I - Phase V - Phase IV. 
Furthermore, the variation of the phase transition critical parameter $\gamma$ under different $U$ reflects the competition between the superconducting paring interaction and non-Hermiticity.

\end{document}